# Traffic Prioritization Mechanisms for Mission and Time Critical Applications in Industrial Internet of Things

Anwar Ahmed Khan[1*], Shama Siddiqui[2], Indrakshi Dey[3]

[1] Millennium Institute of Technology and Entrepreneurship, Pakistan. South East Technological University, Ireland
[2] DHA Suffa University, Pakistan. South East Technological University, Ireland
[3] South East Technological University, Ireland
*Corresponding Author: yrawna@yahoo.com

**Abstract:** Industrial Internet of Things (IIoT) promises to revolutionize industrial operations and productions through utilizing Machine-to-Machine (M2M) communications. Since each node in such environments generates various types of data with diverse service requirements, MAC protocol holds crucial importance to ensure efficient delivery. In this context, simple to complex MAC schemes are found in literature. This paper focuses on evaluating the performance of two major techniques "slot stealing" and "packet fragmentation" for the IIoT; representative protocols SS-MAC and FROG-MAC have been chosen from each category respectively. We conducted realistic simulations for the two protocols using Contiki. Delay and packet loss comparison for SS-MAC and FROG-MAC indicates the superiority of FROG-MAC due to reduction in the waiting time for urgent traffic. Thus, a simple fragmentation scheme could be deployed for efficient scheduling of heterogenous traffic in the industrial environments.

## 1. Introduction

Industrial Internet of Things (IIoT) takes advantage of sensing and communication technology fundamentally to enhance the industrial production. The otherwise dumb industrial equipment is made smart through integration of sensors and actuators which provide valuable data; the equipment autonomously coordinates (referred as M2M communication) for decision-making and reducing human involvement [1]. Some of the major processes facilitated by IIoT include automation of key operations, predictive maintenance, worker safety and well-being, business analytics and utility management. For each of these processes, the nodes generate data of heterogenous types with varying delay tolerance [2]; for instance, the data about emergency shut down versus periodic equipment monitoring would have significantly different service requirements. Therefore, the data requiring urgent delivery needs to be given higher priority as compared to the routine data which may be assigned lower priority.

To deal with the diverse priority requirements in industrial IoT, large number of schemes including routing and MAC protocols have been proposed in the literature [3]. It is to be noted that in addition to meeting the delay requirements, the protocols also need to consider other constraints such as energy, network congestion, buffer overload, and possibility of node/hotspot failures [4]. In this paper, we focus on the MAC protocols targeted specifically to manage urgency requirements of industrial data generated for mission-critical applications.

Initially, the focus of MAC protocols developed for IIoT has been on transmitting data rather than on prioritizing them. With the rapid evolution of industrial IoT applications, the priority mechanisms have been embedded with the MAC schemes. For example, multi-channel MAC has been proposed in to facilitate the mission critical applications [5]; Priority-MAC was proposed for hijacking the transmission slots by urgent data [6]; SS-MAC was also proposed for stealing transmission slots from the nodes with non-critical traffic [7]; FROG-MAC was proposed to fragment the non-urgent data so the urgent data may get an early chance of transmission rather than waiting for the complete transmission of lower priority data [8].

Despite the presence of numerous MAC schemes for heterogenous data, there exist limitations for each. For instance, the schemes based on super frame transmission increase the delay due to the transmission of super frame [9]; the schemes based on slot stealing still cause delay for the prioritized data as it must wait for the completion of ongoing transmission on the channel [10]. Similarly, the schemes which distribute schedules incur delay and additional resources for communicating and maintaining transmission schedules at each node as well as its neighbors [11]. Due to the availability of numerous MAC schemes dealing with prioritized heterogenous traffic, we believe it will be a valuable research contribution to offer a comparison, particularly for the schemes that seem to bring similar outcome for the IIoT scenarios. This work provides a comparative analysis of the above two priority mechanisms, SS-MAC and FROG-MAC for industrial IoT based on realistic experiments. Here, it is to be noted that in the proposal of SS-MAC, no simulations were performed. Hence, this work, in addition to comparing the two schemes, also involves a novel contribution of implementing a full-fledged super-frame-based slot stealing mechanism.

Rest of this paper has been organized as follows: Section II presents a brief overview of the relevant work; Section III details the experimental set-up; Section IV presents the results and evaluation; finally, Section V concludes the paper.

## 2. Relevant Work

In this section, we provide a brief overview of SS-MAC and FROG-MAC which have been selected from the categories of slot-stealing and fragmentation. SS-MAC is developed to enable the time-critical traffic to steal slots from non-critical traffic. It is a synchronous protocol with the conventional mechanism of transmitting super frames so all nodes may align their transmission schedules. As compared to TDMA, SS-MAC guarantees early delivery of time-critical traffic as the non-critical traffic is deferred till the next slot. The timing structure for transmitting critical and non-critical traffic has been illustrated in fig.1 As shown in fig.1 (a), the nodes with non-critical traffic are assigned slots in each cycle, with an Emergency Indication Subslot (EIS) period between two consecutive cycles. In case any node generated a critical packet, it has to wait for EIS to transmit an indication. Upon receiving this indication, the non-critical nodes defer their transmission by one slot.

The operation of critical traffic has been shown in fig 1 (b). Here, it is seen that after transmitting an alert during EIS, the critical nodes have to follow the transmission cycle comprising of three phases namely: Reservation Request Phase (RRP), Deterministic Schedule Phase (DSP), and Data Transmission Phase (DTP). RRP is further divided into subslots to be assigned to each critical node. Once a node has sent an alert during EIS, it then requests the controller for a transmission time slot during RRP. The controller checks the deadline threshold for each packet (in case alert has been received by multiple critical nodes in a given transmission cycle) and allocates the DTP by setting the value of Chanel Allocation Order (CAO). The controller performs scheduling during DSP.

Although SS-MAC has proposed an interesting mechanism for providing early transmission opportunity to the urgent traffic, the critical traffic has to wait for various time periods, such as for EIS and RRF. Moreover, based on CAO, the critical traffic may need to wait for one or more transmission slots before it could get a chance to transmit. This is because if a new node with a low value of threshold sends an EIS, its value of CAO will be lower as compared to the nodes that had been waiting already. Furthermore, it only offers a performance evaluation in terms of mathematical modeling and no simulation/testbed evaluation has been conducted for the proposal.

FROG-MAC [8] is an asynchronous protocol, based on fragmentation of low priority data to facilitate the transmission of higher priority data. The basic timing diagram of this protocol has been illustrated in fig:2. In fig 2(a), node 1 transmits urgent traffic followed by the conventional exchange of RTS/CTS. Since the data is assumed to be of high priority, it is not fragmented and is transmitted as a single unit. On the other hand, fig 2 (b) shows the opposite case where the node has a lower priority data to transmit. In this case, the packet is first fragmented and then transmitted such that there is enough space between consecutive fragments that node with a higher priority data may interrupt the channel. Therefore, FROG-MAC reduces the overall waiting time as it is not required to wait for complete transmission of lower priority packets, which is the case with all the protocols discussed above.

As clear from the above literature review, tens of MAC protocols have been found in the literature which may be deployed for the industrial sensing environments. However, to ensure reliable and timely communication, the comparison of these techniques would be beneficial for enhancing the overall industrial operations. Therefore, in this work we compare the two protocols SS-MAC and FROG-MAC for their delay performance, specifically by generating heterogenous data to represent mission and time critical applications.

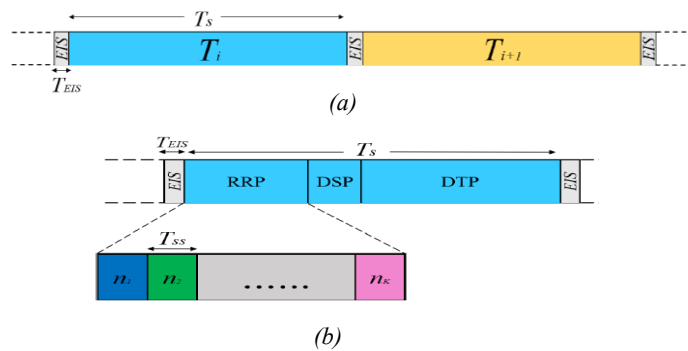

Fig. 1: Transmission Cycles Defined by SS-MAC- (a) For Non-critical Traffic, (b) For Critical Traffic

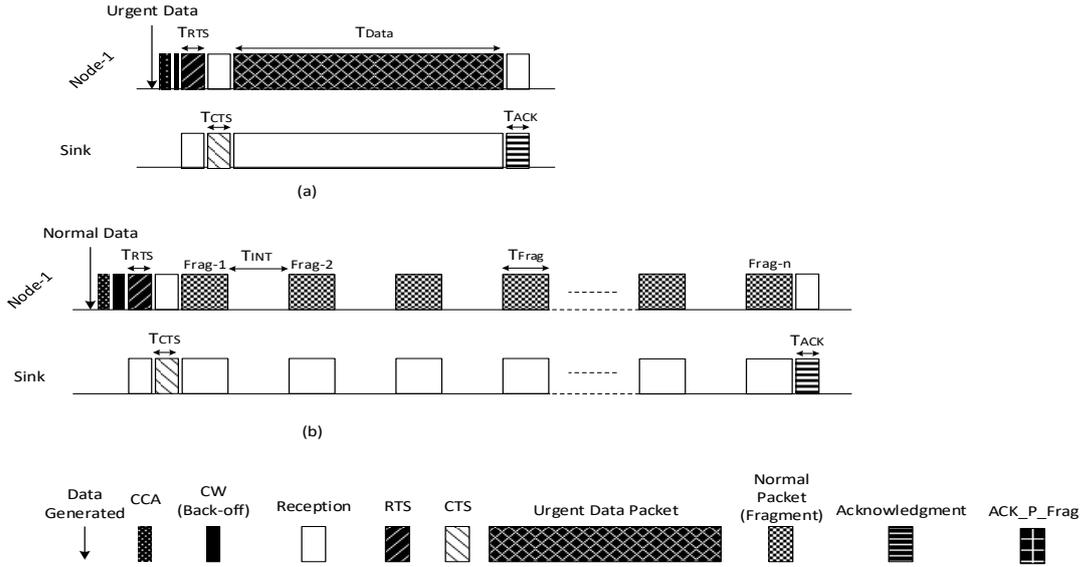

Fig. 2: Basic Operation of FROG-MAC- (a) Transmission of urgent Data, (b) Transmission of Normal Data

## 3. Experimental set-up

Traffic of two priorities was used for each experiment reported in this section to illustrate time critical (urgent) and non-time critical (normal) data, represented by $P_c$ and $P_{nc}$ respectively. To ensure the aspect of heterogeneity and study the packet loss, we considered that urgent data packets dropped if cannot be acknowledged in a certain period. The implementation has been developed at the pin level, so the experiments may be run directly on the nodes. This gives us the confidence that the reported evaluation results will remain valid when practically deployed on the industrial equipment.

A single hop star topology of 20 nodes was considered, where the central node acted as a sink, and number of nodes sending Pc traffic was varied. This is because for various IIoT applications, the number of nodes generating urgent traffic is expected to vary; for example, the nodes reporting fire on the floor will be different as compared to the nodes sending packets to inform about emergency shutdown. Furthermore, since the generation of time critical traffic is expected to be rare as compared to the non-time critical traffic, their generation rate has been kept significantly different. The major simulation parameters used in the experiments have been presented in table 1.

Table 1: Simulation Settings for Comparing SS-MAC and FROG-MAC

| Simulation Parameter | | Simulation Settings |
|---|---|---|
| Simulation Duration | | 5000 Sec |
| Total Number of Nodes | | 20 |
| Number of Transmitting nodes | | Variable |
| Message Generation Interval of | Urgent/TC1 Traffic | 2 min |
| | Normal/TC3 Traffic | 10 sec |
| Data Packet Length | | 34 Bytes |
| Fragment Size for FROG-MAC | | Varying (2 to 32) |

## 4. Results and Evaluation

Fig 3 illustrates the average delay performance comparison for SS-MAC and FROG-MAC for the critical/urgent traffic. Here, we define average delay as an average of delays faced by each urgent packet between its generation and reception. With the increasing number of nodes transmitting urgent data, the delay for SS-MAC is seen to increase significantly, whereas the impact is very low for FROG-MAC. In SS-MAC, each node will get a transmission opportunity after subsequent EIS and based on its subslot's position in the RRF; on the other hand, in FROG-MAC, each node will contend with others using the conventional back-off and contention window mechanisms. Therefore, if an urgent packet is generated earlier by a node, it will be transmitted earlier.

As seen from the fig 3., there is a significant difference in delay performance observed for FROG-MAC and SS-MAC. Firstly, there is a fundamental difference of synchronization between the two protocols; SS-MAC is a synchronous scheme that requires periodic transmission of super frames, so all the nodes remain aware about the transmission cycles of their neighbors. On the other hand, FROG-MAC not only reduces the resource consumption but also improves the delay performance by taking an asynchronous approach. As a result, difference in delay is observed as the nodes operating on FROG-MAC do not need to transmit or process the super frames.

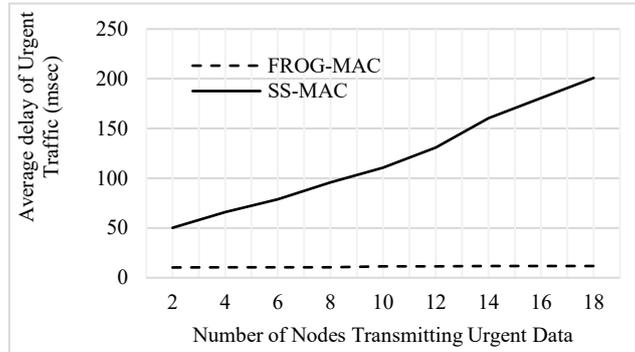

Fig. 3: Delay comparison for varying transmitting nodes

The major rationale behind improved delay of FROG-MAC is that Pc traffic just has to wait for a short fragment's transmission, rather than waiting for EIS and RRF in SS-MAC. As seen from fig 4, the transmission cycle of non-critical traffic comprises of entire transmission slot that a critical node may not interrupt. Although due to the introduction of EIS between the transmission slots, the critical traffic does get a chance to defer the non-critical traffic, it still has to wait for the complete transmission of this data; the same issue of not being able to interrupt an ongoing transmission on the channel motivated us to introduce the concept of packet fragmentation. Even after waiting for a complete transmission slot, the critical node does not get a chance to transmit its data, but it can only transmit an indication about emergency.

In addition to the need for waiting for EIS to transmit an indication, the critical node now needs to wait for its turn for transmitting the data. Based on the deadline information, the node will be allocated a CAO value. Although this approach will ensure that the packet with the highest level of criticality gets transmitted first, the delay of a typical packet will increase. On the other hand, FROG-MAC deals with the problem of concurrent transmission of data by multiple critical nodes by utilizing the conventional contention window mechanism. The nodes which generate data first will have low values of contention window as compared to the nodes that request for the medium later in the sequence.

Fig 4 further illustrates the comparison of delay of the two protocols for increasing number of transmitting nodes, while also gradually increasing the fragment size. As previously discussed, the increase in fragment size is expected to increase the delay for urgent traffic because it must wait for longer. As seen from the fig 4, there is negligible impact on delay of FROG-MAC for the increasing fragment size, however, the delay of SS-MAC remains higher for all fragment sizes. This is again due to various waiting timers involved in SS-MAC which does not let the critical traffic to get transmitted immediately after getting access to the channel, as is the case with FROG-MAC.

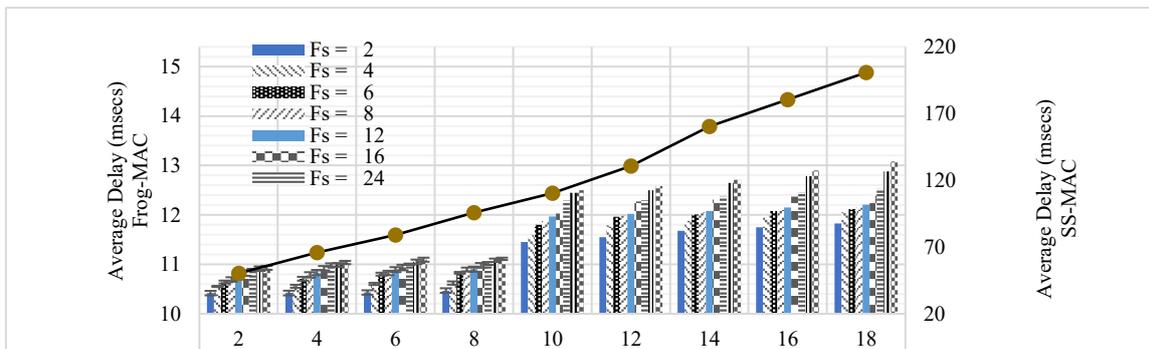

Fig. 4: Delay comparison for varying transmitting nodes and fragment size

Finally, Figure 5 compares the packet loss for FROG-MAC and SS-MAC for varying the number of nodes that transmit urgent data. Here, it can be seen that the packet loss increases with the increasing urgent traffic for both protocols, however, the increase is more rapid and significant for SS-MAC. Just like the delay observation in previous figures, the packet loss is also on a higher side for SS-MAC due to the requirement of higher wait for the urgent traffic. Therefore, the approach of fragmentation appears to be better than slot stealing in terms of packet loss as well.

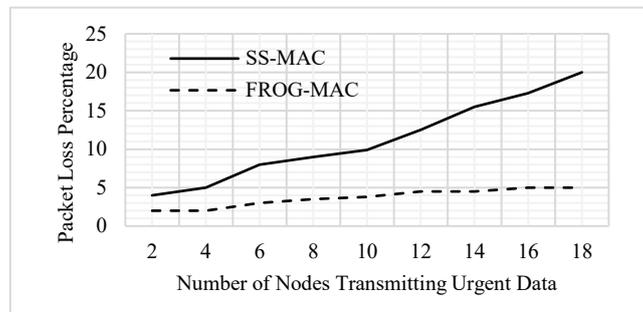

Fig. 5: Packet Loss comparison for varying transmitting nodes

## 5. Conclusion

This paper provided a performance comparison of two common MAC schemes often used in heterogenous traffic environments. SS-MAC has been taken to represent slot stealing protocols, whereas FROG-MAC was selected from the category of fragmentation-based protocols. It has been observed that mission critical traffic faces lesser delay and packet loss for FROG-MAC as compared to the SS-MAC due to its unique design of reducing waiting time by fragmenting the non-critical traffic. Therefore, the fragmentation scheme appears to be better both in terms of latency and reliability.

In future, we plan to enhance the study by evaluating the protocols for industrial scenarios with dynamically changing priorities. Furthermore, the comparisons of various priority-based schemes is planned to be performed on industrial testbeds such as Electric Power Intelligent Control (EPIC).

## 6. Acknowledgment

This contribution is supported by HORIZON-MSCA 2022-SE-01-01 project COALESCE under Grant Number 10113073.